\documentstyle[12pt]{article}
\newlength{\extraspace}
\newlength{\extraspaces}
\newcommand{\be}{\begin{equation}
\addtolength{\abovedisplayskip}{\extraspaces}
\addtolength{\belowdisplayskip}{\extraspaces}
\addtolength{\abovedisplayshortskip}{\extraspace}
\addtolength{\belowdisplayshortskip}{\extraspace}}
\newcommand{\ee}{\end{equation}}

\newcommand{\ba}{\begin{eqnarray}
\addtolength{\abovedisplayskip}{\extraspaces}
\addtolength{\belowdisplayskip}{\extraspaces}
\addtolength{\abovedisplayshortskip}{\extraspace}
\addtolength{\belowdisplayshortskip}{\extraspace}}
\newcommand{\ea}{\end{eqnarray}}

\newcommand{\newsection}[1]{
\vspace{15mm}
\pagebreak[3]
\addtocounter{section}{1}
\setcounter{equation}{0}
\setcounter{subsection}{0}
\setcounter{footnote}{0}
\begin{center}
{\large \thesection. #1}
\end{center}
\nopagebreak
\medskip
\nopagebreak}

\newcommand{\newsubsection}[1]{
\vspace{1cm}
\pagebreak[3]

\addtocounter{subsection}{1}
\noindent{ \sc \thesubsection. #1}
\nopagebreak
\vspace{2mm}
\nopagebreak}

\def\fr#1#2{{#1\over #2}}
\def\du#1{\tilde #1}
\def\mat#1#2#3#4{{\left(\matrix{#1&#2\cr#3&#4\cr}\right)}}

\def\f{\phi}
\def\F{\Phi}
\def\G{\Gamma}

\def\th{\theta}
\def\td{\du\th}

\def\a{\alpha}

\def\l{\lambda}

\def\S{\Sigma}

\def\pa{\partial}
\def\pab{\bar\pa}

\def\ie{{\it i.e.},\ }

\def\cf{{\it cf.}\ }
\def\ind{{1\over 2\pi}\int\!\! d^2\! z\,}

\def\AB{{\bar A}}
\def\JB{{\bar J}}
\def\q{\, , \qquad}
\def\sq{\, , \quad}
\def\R{{\bf R}}
\def\Z{{\bf Z}}
\def\[{\Bigl[}
\def\]{\Bigr]}
\def\gl{\G^1_{ai}}
\def\gz{\G^2_{ia}}
\def\gab{\G_{ab}}
\def\ra{\rightarrow}

\def\t{tanh^2x}

\def\sh{sinh^2x}
\def\ch{cosh^2x}

\catcode`@=11           
\dimendef\dimen@=0
\def\ialign{\everycr{}\tabskip\z@skip\halign} 
\def\m@th{\mathsurround=\z@}
\def\openup{\afterassignment\@penup\dimen@=}
\def\@penup{\advance\lineskip\dimen@
  \advance\baselineskip\dimen@ \advance\lineskiplimit\dimen@}
\def\eqalign#1{\null\,\vcenter{\openup\jot\m@th
  \ialign{\strut\hfil$\displaystyle{##}$&$\displaystyle{{}##}$\hfil
\crcr#1\crcr}}\,}
\catcode`@=12           

\def\,{\kern .16667em}  

\begin{document}
\addtolength{\baselineskip}{.7mm}
\thispagestyle{empty}
\begin{flushright}
{\sc IASSNS-HEP}-91/84\\ {\sc ITP-SB}-91-67\\ December 1991
\end{flushright}
\vspace{1cm}

\centerline{\Large \bf Generalized Duality in Curved String-Backgrounds}
\vskip 1in

\centerline{\large Amit Giveon\footnote{Email: giveon@iassns.bitnet}
 and Martin Ro\v cek\footnote{ Permanent address: ITP, SUNY at Stony
Brook, Stony Brook NY 11794-3840. \\ Email:
rocek@dirac.physics.sunysb.edu}}
\vspace{1.5cm}

\centerline{\it School of Natural Sciences}
\centerline{\it Institute for Advanced Study}
\centerline{\it Princeton, NJ 08540}

\vspace{1.5cm}

\centerline{\bf Abstract}
\vspace{.2cm}

\noindent
\parbox{15cm}{The elements of $O(d,d,\Z)$ are shown to be
discrete symmetries of the space of curved string backgrounds that are
independent of $d$ coordinates.  The explicit action of the symmetries
on the backgrounds is described.  Particular attention is paid to the
dilaton transformation.  Such symmetries identify different
cosmological solutions and other (possibly) singular backgrounds; for
example, it is shown that a compact black string is dual to a charged
black hole. The extension to the heterotic string is discussed.}
\vfill
\eject


\newsection{Introduction}

\noindent
In string theory, conformal field theories (CFT's) correspond to
classical vacua (for a review, see \cite{gsw}).  In general, a CFT can
be deformed by truly marginal operators.  The space of couplings to
these operators is a connected subspace of the space of string vacua.
The space of all string vacua has many such components (which may
approach the same boundaries).  Many tractable cases involve CFT's
that can be described by a sigma-model action, which can be thought of
as a string moving in a metric, antisymmetric tensor, and dilaton
background.  In these theories, some marginal operators generate
deformations of the background.

Different string theory backgrounds may correspond to the same CFT.
The simplest example of this phenomenon is the $R\ra 1/R$ duality for
a free boson compactified to a circle \cite{rdual}.  For
$d$-dimensional toroidal backgrounds, this duality generalizes to a
discrete symmetry group isomorphic to $O(d,d,\Z )$ (or $O(d,d+16,\Z )$
for the heterotic string)
\cite{grv,sw,gmr}.

In this paper, we discuss the discrete symmetry group acting on the
space of curved $D$-dimensional backgrounds that are independent of
$d$ coordinates $(d<D)$.  Such backgrounds include many explicitly
known string vacua: black holes \cite{bh}, $p$-branes \cite{branes},
cosmological solutions \cite{cosmo,tv}, etc.

The discrete symmetries identify vacua with geometries that in general
are radically different.  This characteristic stringy property is a
consequence of the possibility that strings can wind around compact
coordinates.  The geometries related by the discrete symmetries differ
in their identification of the (local) momentum modes and the
(non-local) winding modes.  Thus, to an observer composed of, {\it
e.g.,\/} momentum modes, geometries mathematically equivalent as CFT's
will appear physically distinct.

The discrete symmetries in general act on the dilaton; in the flat
case, the constant dilaton is transformed to a new constant such that
the string coupling remains invariant \cite{gv,gmr}.  In the curved
case, the dilaton transforms analogously \cite{betal}.  The dilaton
transformation plays an important role in the understanding of the
propagation of a string in the $D=2$ black hole geometry \cite{gdvv},
and in the understanding of duality invariant cosmological solutions
\cite{tv}.

The structure of the moduli space that the discrete symmetries act on
is not known in general.  In the flat ($D$-dimensional) case, the
moduli space is locally {isomorphic} to $O(D,D,\R )/(O(D,\R ))^2$
\cite{nar}.  Studies of low energy effective actions suggest that
there is a moduli subspace of backgrounds independent of $d$
coordinates isomorphic to $O(d,d,\R )/G$ \cite{gmr2,gpo,ven}, where
$G$ is at least the diagonal subgroup $O(d,\R )$ of $(O(d,\R ))^2$
(the maximal compact subgroup of $O(d,d,\R )$).  This is consistent
with results from string field theory \cite{sen}.  This moduli
subspace is only correct when the $d$ coordinates have the topology of
a torus.  Though in general the local structure of the moduli space is
unknown, we can still identify a discrete symmetry group that acts on
it.

The main result of this paper is that there is a discrete symmetry
group transforming curved $D$-dimensional backgrounds independent of
$d$ coordinates $(d<D)$, which is isomorphic to $O(d,d,\Z )$.  This
group is naturally embedded in $O(D,D,\Z )$, and acts on the
$D$-dimensional (metric + antisymmetric tensor) background by
fractional linear transformations, together with a dilaton
transformation that preserves the string coupling.  Though the
explicit transformations are valid only to leading order in the
inverse string tension $\a'$, we generalize the result of \cite{rv} to
argue that the symmetry survives to all orders.

The basic example of a nontrivial CFT that has two curved spacetime
interpretations related by duality is the $D=2$ black hole
\cite{gdvv}.  More general discussions of such duality have appeared
in \cite{betal,ketal}.

The paper is organized as follows: In section 2, we generalize the
result of \cite{rv} and construct general (conformal) curved
$D$-dimensional backgrounds that are independent of $d$ coordinates as
abelian quotients
\cite{gk} of CFT's with $(D+d)$-dimensional backgrounds.
In section 3, we use the construction of section 2 to find the
discrete symmetries of the space of these $D$-dimensional backgrounds.
In section 4, we explore the group structure of these symmetries, and
find a group isomorphic to $O(d,d,\Z )$, as well as a simple
expression for its action on the backgrounds. In section 5, we focus
on the dilaton and its transformations.  In section 6, based on
analogies with the flat case, we conjecture how these results extend
to the heterotic string.  In section 7, we study two examples: duality
between compact black strings and charged black holes in the bosonic
string, and duality between neutral and charged black holes in the
heterotic string.  In section 8, we close with a few comments and
discuss related open problems.

\newpage

\newsection{Curved target spaces independent of $d$ coordinates as quotients}

\noindent
In this section we construct a general (conformal) curved background
in $D$ dimensions that is independent of $d$ coordinates as an abelian
quotient of a CFT with a $(D+d)$-dimensional background.  Generalizing
the result of \cite{rv}, we start with a CFT with $d$ abelian left
handed currents $J^i$ and right handed currents $\JB^i$.  The action
is
\be
\label{Dd}
\eqalign{
S_{D+d}=&\, S_1+S_a+S[x] \, ,\cr S_1=&\, \ind \[\pa\th_1^i\pab\th_1^i
+ \pa\th_2^i\pab\th_2^i+ 2\S_{ij}(x)\pa\th_2^i\pab\th_1^j +\gl (x)\pa
x^a\pab\th_1^i + \gz (x)\pa\th_2^i\pab x^a\]\cr S_a=&\,
\ind\[\pa\th_1^i\pab\th_2^i - \pa\th_2^i\pab\th_1^i\]\cr S[x]=&
\,\ind\[\gab (x)
\pa x^a \pab x^ b -\fr14\F(x) R^{(2)} \]\, ,}
\ee
where $i,j = 1 ,\dots ,d$ and $a, b = d+1 ,\dots ,D$, and $\S_{ij},\gl
,\gz , \gab$ are components of arbitrary $x$-dependent matrices, such
that, together with the dilaton $\F$, the theory described by the
action $S_{D+d}$ is conformal.

The antisymmetric term $S_a$ is (locally) a total derivative, and
therefore may give only topological contributions, depending on the
periodicity of the coordinates $\th$.\footnote{This term was omitted
in \cite{rv}; it is needed for gauge invariance of the gauged action,
see below.} To specify the periodicity, we define
\begin{equation}
\label{th}
\th^i=\th_2^i-\th_1^i \q \du\th^i=\th_1^i+\th_2^i \, ,
\end{equation}
such that
\begin{equation}
\label{per}
\th^i\equiv\th^i+2\pi \q \du\th^i\equiv\du\th^i+2\pi \, .
\end{equation}
In these coordinates, $S_a$ becomes
\be
\label{Sa}
S_a = \ind \fr12 \[\pa\du\th^i\pab\th^i -\pa\th^i\pab\du\th^i\]
\, ,
\ee
which takes half-integer values, and therefore contributes to the
path-integral.

The action $S_{D+d}$ (\ref{Dd}) is invariant under the $U(1)_L^d\times
U(1)_R^d$ affine symmetry generated by currents
\be
\label{cur}
\eqalign{
J^i\, =\, &\pa\th_1^i+\S_{ji}\pa\th_2^j +\fr12\G^1_{ai}\pa x^a\cr
\, =\, &\fr12\[-(I-\S )_{ji}\pa\th^j +(I+\S )_{ji}\pa\td^j +\G^1_{ai}\pa
x^a\]\,
    ,
\cr
\JB^i\, =\, &\pab\th_2^i+\S_{ij}\pab\th_1^j +\fr12\G^2_{ia}\pab x^a\cr
\, =\, &\fr12\[(I-\S )_{ij}\pab\th^j +(I+\S )_{ij}\pab\td^j +
\G^2_{ia}\pab x^a\]\, .\cr
}
\ee
We choose to gauge the $d$ anomaly-free axial combinations of the
symmetries (\ref{cur});\footnote{In \cite{rv}, the designation of
axial {\it vs.\/} vector was interchanged.} other options are
generated by discrete symmetries discussed later.  The gauged action
is \cite{gk}
\begin{equation}
\label{gauge}
S_{gauged}=S_{D+d}+\ind\[A^i\JB^i+\AB^iJ^i+
\fr12A^i\AB^j(I+\S )_{ij}\]\, .
\end{equation}
The antisymmetric term $S_a$ (\ref{Sa}) is needed to insure gauge
invariance under large gauge transformations.\footnote{Indeed, with
the choice of $S_a$ above, minimal coupling $\pa_\a \du\th^i \ra
\pa_\a\du\th^i+A^i_\a$ gives the correct gauged model.} Integrating
out the gauge fields $A^i,\AB^i$ gives:
\be
\label{D}
\eqalign{
S_D\, =\, &\ind \[{\cal E}_{IJ}(x)\pa X^I \pab X^J -\fr14\f (x)
R^{(2)}\]\cr
\, =\, &\ind\[E_{ij}(x)\pa\th^i\pab\th^j+F^2_{ia}(x)\pa\th^i\pab x^a+
F^1_{ai}(x)\pa x^a\pab\th^i \cr &\qquad\qquad\qquad\qquad +
F_{ab}(x)\pa x^a \pab x^b -\fr14 \f(x) R^{(2)}\]\, ,}
\ee
where $$
\{ X^I\}_{I=1\dots D}=\{\th^i,x^a\}_{i=1\dots d,a=d+1\dots D}
$$ and
\be
\label{calG}
{\cal E}_{IJ} = {\cal G}_{IJ}+{\cal B}_{IJ}
=\left(\matrix{E_{ij}&F^2_{ib}\cr F^1_{aj}&F_{ab}\cr}\right)\, .
\ee
We have split $\cal E$ into its symmetric and antisymmetric parts
$\cal G$ and $\cal B$.  The components are
\be
\label{ij}
E_{ij}=(I-\S )_{ik}(I+\S )^{-1}_{kj} \, ,
\ee
and
\be
\label{F}
F^2_{ia}=(I+\S )^{-1}_{ij}\G^2_{ja}\sq F^1_{ai}=-\G^1_{aj}(I+\S
)^{-1}_{ji}
\sq F_{ab}=\gab-\fr12\gl (I+\S )^{-1}_{ij} \G^2_{jb} \, .
\ee
This $D$-dimensional target space background is independent of the $d$
coordinates $\th^i$.  Since the relations (\ref{ij},\ref{F}) can be
inverted to solve for $(\S , \G )$ in terms of $(E,F)$, it is the most
general such background.  Following the reasoning of \cite{rv}, if the
original model $S_{D+d}$ (\ref{Dd}) is conformal, then $S_D$ is
conformal to one loop order with\footnote{However, as noted in
\cite{rv}, higher order corrections to the background that give an
{\it exact\/} CFT with a $D$-dimensional background exist; these
corrections come from the integration measure.}
\begin{equation}
\label{f}
\f=\F+\ln det (I+\S )\, .
\end{equation}
This relation is also invertible, and thus a $D$-dimensional field
theory with a background that is independent of $d$ coordinates can be
described as a quotient of a $(D+d)$-dimensional field theory with $d$
chiral currents.  If the $D$-dimensional theory is conformally
invariant, then the $(D+d)$-dimensional theory is as well \cite{rv},
and hence {\it any\/} CFT with a background that is independent of $d$
coordinates can be described as a quotient of a CFT with a
$(D+d)$-dimensional background.

This construction will allow us to understand the discrete symmetries
of the moduli space of string vacua in backgrounds that are
independent of $d$ coordinates.

\newsection{Discrete Symmetries}

\noindent
Different string theory backgrounds may describe the same conformal
field theory.  Here we study discrete symmetries that relate different
but equivalent backgrounds $(E(x),F(x),\f (x))$ (\ref{D}).  We first
discuss transformations of $(E(x),F(x),\f (x))$ that follow from
manifest symmetries of the action $S_{D+d}$ (\ref{Dd}).  We then
combine them with transformations that are manifest symmetries of
$S_D$ (\ref{D}) to find a discrete symmetry group isomorphic to
$O(d,d,\Z )$.

The action $S_{D+d}$ is invariant under the transformations
\be
\label{OxO}
\th_1 \ra O_1\th_1 \q \th_2 \ra O_2\th_2 \q O_{1,2}\in O(d,\Z ) \, ,
\ee
together with
\be
\label{OSO}
\S\ra O_2\S O_1^t\q\G^1\ra\G^1O_1^t\q\G^2\ra O_2\G^2\q\G\ra\G
\ee
such that
\be
\label{cond}
\fr12 ( O_1 \pm O_2 )_{ij} \in \Z \, .
\ee
Here $O(d,\Z )$ is the group of matrices $O$ with integer entries
satisfying $OO^t=I$.

These symmetries can be found as follows: The action $S_{D+d}$ is
invariant up to total derivatives under $O(d,\R )\times O(d,\R )$
acting on $(\th_1 ,\th_2)$ as in (\ref{OxO}), together with
(\ref{OSO}) for the backgrounds.  The periodic coordinates $\th , \td$
(\ref{th}) transform as:
\be
\label{mat}
\left(\matrix{\th\cr\td}\right)\ra
\fr12\left(\matrix{O_1+O_2&O_1-O_2\cr O_1-O_2&O_1+O_2\cr}\right)
\left(\matrix{\th\cr\td}\right)
\ee
To preserve the periodicities of $\th ,\du\th$ (\ref{per}), the
condition (\ref{cond}) must be satisfied.  In particular, this implies
$O_{1,2}\in O(d,\Z )$.

The total derivative comes from the transformation of $S_a$
(\ref{Sa}), and is
\be
S_a[O_1\th_1,O_2\th_2]-S_a[\th_1,\th_2] = \ind \[ M_{ij}(\pa\td^i
\pab\td^j-\pa\th^i
\pab\th^j )+N_{ij}(\pa\th^i\pab\td^j-\pa\td^i\pab\th^j)\] \, ,
\ee
where
\be
M=\fr14(O_1^t-O_2^t)(O_1+O_2) \q N=\fr14(O_1^t-O_2^t)(O_1-O_2) \, .
\ee
The condition (\ref{cond}) implies that $M_{ij},N_{ij}\in \Z$, and
hence the total derivative is an integer, and does not contribute to
the path integral.  This concludes the proof that
(\ref{OxO},\ref{OSO}) are symmetries of the action $S_{D+d}$.

The transformations of the background (\ref{OSO}) induce
transformations of the background $(E(x),F(x),\f (x))$
(\ref{ij},\ref{F},\ref{f}) in $S_D$ (\ref{D}):
\be
\label{tre}
\quad \,\, \, E\quad\ra\quad E'\quad =
\quad (I-O_2\S O_1^t)(I+O_2\S O_1^t)^{-1}
\ee
\be
\label{trf}
\matrix{F^1&\ra &F^1{}'&=& -\G^1(O_1+O_2\S )^{-1}\cr
&&&&\cr F^2&\ra &F^2{}'&=& (O^t_2+\S O^t_1)^{-1}\G^2\cr &&&&\cr F&\ra&
F'&=&\G-\fr12\G^1(O^t_2O_1+\S )^{-1}\G^2\cr &&&&\cr}
\ee
\be
\label{trd}
\f\ra\f' =\F+\ln det(I+O_2\S O^t_1 ) \, .
\ee
These transformations can be rewritten as
\be
\label{Etrans}
E'=\[(O_1+O_2)E+(O_1-O_2)\]\[(O_1-O_2)E+(O_1+O_2)\]^{-1}
\ee
\be
\label{Ftrans}
\eqalign{
F^1{}'\, =\, &2F^1\[(O_1-O_2)E +(O_1+O_2)\]^{-1}\cr &\cr F^2{}'\, =\,
&\fr12 \[(O_1+O_2)-E'(O_1-O_2)\] F^2\cr &\cr F'\, =\,
&F-F^1\[(O_1-O_2)E +(O_1+O_2)\]^{-1}(O_1-O_2)F^2\cr &\cr}
\ee
\be
\label{dtrans}
\f'=\f+\fr12\ln \[\fr{det{\cal G}}{det{\cal G}'}\]
=\f+\fr12\ln \[\fr{det G}{det G'}\]
\ee
\vskip .15in
\noindent
where $\cal G$ is the background metric as defined in (\ref{calG}) and
$G$ is the symmetric part of $E$ (\ref{ij}).  We discuss the dilaton
transformation (\ref{dtrans}) in detail in section 5.

The transformations (\ref{OxO},\ref{OSO}) induce a non-trivial action
on the currents (\ref{cur}), and hence $S'_D$ ($S_D$ with a
transformed background $(E',F',\f')$) is derived from $S_{D+d}$
(\ref{Dd}) by a different quotient.  For example, some symmetries
simply change the sign of a $J^i$ without changing $\JB^i$; this
corresponds to a {\it vector\/} gauging (as opposed to an axial
gauging) of the $i$'th $U(1)$.

We now consider the additional transformations that are manifest
symmetries of the action $S_D$ itself.  The first are integer
``$\Theta$"-parameters that shift $E$: $$ E_{ij}\ra
E_{ij}+\Theta_{ij}\q \Theta_{ij}=-\Theta_{ji}\in\Z $$
\be
\label{Th}
F^1\ra F^1 \q F^2\ra F^2 \q F\ra F \, .
\ee
We refer to the group generated by these transformations as $\Theta
(\Z )$; these are obviously symmetries, as they shift the action $S_D$
(\ref{D}) by an integer and thus do not contribute to the path
integral.

The second type of transformations are given by homogeneous
transformations of $E,F^1,F^2$ under $A\in Gl(d,\Z )$:
\be
\label{gl}
E\ra A^t E A \q F^2\ra A^t F^2\q F^1\ra F^1 A\q F\ra F\, .
\ee
These are obviously symmetries of the theory described by $S_D$, as
they generate a change of basis in the space of $\th$'s that preserves
their periodicities.

Neither the $\Theta (\Z )$ nor the $Gl(d,\Z )$ transformations affect
the dilaton, since they do not change the integration measure of the
path integral; consequently, they are symmetries of the exact CFT's,
and receive no higher order corrections.

The group generated by all the symmetries discussed is isomorphic to
$O(d,d ,\Z )$.  A natural embedding of $O(d,d,\Z )$ in $O(D,D,\Z )$
acts on the background $\cal E$ by fractional linear transformations,
as explained in the next section.

\newsection{The action of $O(d,d,\Z )$}

\noindent
We begin by establishing our notation following \cite{gmr2}.  The
group $O(d,d,\R )$ can be represented as $2d\times 2d$-dimensional
matrices $g$ preserving the bilinear form $J$:
\be
\label{g}
g=\left(\matrix{a&b\cr c&d\cr}\right) \q J=\left(\matrix{0&I\cr
I&0\cr}\right)\, ,
\ee
where $a,b,c,d,I$ are $d\times d$-dimensional matrices, and
\be
\label{abcd}
g^tJg=J\quad \Rightarrow \quad a^tc+c^ta=0\sq b^td+d^tb=0\sq
a^td+c^tb=I\, .
\ee
This has an obvious embedding in $O(D,D,\R )$ as
\be
\hat g=\left(\matrix{\hat a&\hat b\cr \hat c&\hat d\cr}\right)
\ee
where $\hat a, \hat b, \hat c, \hat d$ are $D\times D$-dimensional
matrices of the form
\be
\hat a={\left(\matrix{a&0\cr 0&I}\right)}\sq
\hat b={\left(\matrix{b&0\cr 0&0}\right)}\sq
\hat c={\left(\matrix{c&0\cr 0&0}\right)}\sq
\hat d={\left(\matrix{d&0\cr 0&I}\right)}
\ee
(here $I$ is the $(D-d)\times (D-d)$-dimensional identity matrix).

We define the action of $\hat g$ on $\cal E$ by fractional linear
transformations:
\be
\label{ghat}
\eqalign{
\hat g({\cal E}) =& {\cal E}'=
(\hat a {\cal E} + \hat b)(\hat c {\cal E} +\hat d)^{-1}
\cr &\cr
=&\left(\matrix{E'&(a-E'c)F^2\cr
F^1(cE+d)^{-1}&F-F^1(cE+d)^{-1}cF^2\cr}
\right)}
\ee
where
\be
\label{tE}
E'=(aE+b)(cE+d)^{-1}
\ee
is a fractional linear transformation of $E$ under $O(d,d)$.

The group $O(d,d)$ is generated by \cite{gmr2}:

\noindent
$Gl(d)$:
\be
\label{GLd}
\left(\matrix{a&b\cr c&d\cr}\right)=
\left(\matrix{A^t&0\cr 0&A^{-1}\cr}\right) \quad {\rm s.t.}
\quad A\in Gl(d) \, .
\ee
$\Theta$:
\be
\label{THe}
\left(\matrix{a&b\cr c&d\cr}\right)=
\left(\matrix{I&\Theta\cr 0&I\cr}\right) \quad {\rm s.t.}\quad
\Theta=-\Theta^t \, .
\ee
Factorized duality:
\be
\label{FAd}
\left(\matrix{a&b\cr c&d\cr}\right)=
\left(\matrix{I-e_1&e_1\cr e_1&I-e_1\cr}\right) \quad {\rm s.t.}\quad
e_1=diag(1,0,\dots ,0 ) \, .
\ee
The maximal compact subgroup of $O(d,d)$ is $O(d)\times O(d)$ embedded
as
\be
\label{Odd}
\left(\matrix{a&b\cr c&d\cr}\right)=
\fr12\left(\matrix{o_1+o_2&o_1-o_2\cr o_1-o_2&o_1+o_2\cr}\right)
\quad {\rm s.t.}\quad o_{1,2}\in O(d) \, .
\ee
This subgroup includes factorized duality (\ref{FAd}).

We now turn from definitions to the actual symmetries of the CFT.
These form an $O(d,d,\Z )$ discrete subgroup of $O(d,d,\R )$ that acts
on the background as above.  The elements of the subgroup $O(d,d,\Z )$
are given by matrices $g$ of the form (\ref{g},\ref{abcd}) with {\it
integer\/} entries.

Just as in the continuous case, the discrete group is generated by
$Gl(d,\Z )$, $\Theta (\Z )$, and factorized duality; these are given
by (\ref{GLd},\ref{THe},\ref{FAd}) with integer entries.  The subgroup
$O(d,\Z )\times O(d,\Z )$ is given by the matrices (\ref{Odd}), again
with integer entries.  Clearly, the $O(d,\Z )\times O(d,\Z )$
symmetries (\ref{Etrans},\ref{Ftrans}), the $\Theta (\Z )$ symmetries
(\ref{Th}), and the $Gl(d,\Z )$ symmetries (\ref{gl}) that we found in
the previous section act on the background by the $O(d,d,\Z )\subset
O(D,D,\Z )$ fractional linear transformations (\ref{ghat}) with the
matrices $a,b,c,d$ given by (\ref{Odd}, with $o_{1,2}=O_{1,2}$),
(\ref{THe}, with $\Theta\in\Z$), and (\ref{GLd}, with $A\in Gl(d,\Z
)$), respectively.

These results are compatible with the known discrete symmetries of the
space of flat $D$-dimensional toroidal backgrounds \cite{grv,sw,gmr}.
In that case, the $O(d,d,\Z )$ symmetries described above are simply a
subgroup of the full $O(D,D,\Z )$ symmetry group acting as in
(\ref{ghat}), for {\it any\/} $\hat g\in O(D,D,\Z )$.  For curved
$D$-dimensional backgrounds, we expect that some large symmetry group
acts (analogous to $O(D,D,\Z )$); here we have described the $O(d,d,\Z
)$ subgroup that is associated with a $d$-dimensional toroidal
isometry of the background.

In the flat case, the fractional linear transformation is an exact map
between equivalent backgrounds; in the curved case, in general one
expects higher order corrections to the transformed background
\cite{betal}.  For $\Theta (\Z )$ and $Gl(d,\Z )$, the transformations
are exact; however, the factorized duality receives corrections from
the path-integral measure.  Nevertheless, because the transformation
is exact in the ($D+d$)-dimensional model, we know that
non-perturbative correction must exist such that factorized duality is
exact.\footnote{$N=4$ supersymmetry can protect duality
transformations, and there are examples where the one loop
transformations are exact even in curved backgrounds.}

We close this section with a general remark.  The group $O(d,d)$ has
two disconnected components (of the generators given in
(\ref{GLd},\ref{THe},\ref{FAd}), only factorized duality (\ref{FAd})
has $det=-1$, and hence is not connected to the identity).  Therefore
one expects that in general the submoduli space generated by $O(d,d)$
can be disconnected.  In the flat case, it happens that the $O(d,d)$
acts on the moduli space by a double covering, and as a result, the
moduli space is connected.  However, in general one gets two
disconnected components of backgrounds with different topologies: for
example, in the $D=2$, $d=1$ curved case, the two-dimensional
black-hole and its dual are in two disconnected components mapped into
each other by $O(1,1)$.  The duality transformation identifies the two
components as conformal field theories. This is a general feature:
factorized duality (\ref{FAd}) maps one component of the moduli space
of backgrounds to the other.\footnote{Factorized duality is similar to
mirror symmetry \cite{gpcan}, which also identifies two (possibly)
disconnected components of moduli space corresponding to backgrounds
with different topologies. For $N=2$ toroidal (orbifold) backgrounds,
mirror symmetry and factorized duality are identical~\cite{gs}.}

\newsection{The dilaton}

\noindent
To complete the previous discussion of the transformation of the
background under the $O(d,d,\Z )$ symmetries, we consider the
transformation of the dilaton.  This is summarized in eq.
(\ref{dtrans}), which we derive by proving a theorem: The quantity
\be
\label{them}
\du\f=\f+\fr12\ln det{\cal G}
\ee
is invariant under $O(d,d,\Z )$ transformations.  This implies
\be
\label{tdil}
\f'=\f+\fr12\ln\Big( \fr{det{\cal G}}{det{\cal G'}}\Big)\, .
\ee
The second equality in (\ref{dtrans}) follows from the proof of
(\ref{them}) given below.

We begin with the identity
\be
\left(\matrix{G_{ij}&G_{ib}\cr G_{aj}&G_{ab}}\right)=
\left(\matrix{G_{ik}&0\cr G_{ak}&I_{ac}}\right)
\left(\matrix{I_{kj}&(G^{-1})_{kl}G_{lb}\cr 0&G_{cb}
-G_{ck}(G^{-1})_{kl}G_{lb}}\right)
\ee
which implies
\be
\label{det}
det({\cal G})=det(G_{ij})det(G_{ab}-G_{ak}(G^{-1})_{kl}G_{lb})\, .
\ee
We next prove that the following two quantities are separately
invariant under $O(d,d,\Z )$ transformations:
\be
\label{lem}
{\rm~the~invariant~fiber~dilaton:~~} \hat\f=\f+\fr12\ln det(G_{ij})\,
,
\ee
\be
\label{quot}
{\rm~the~quotient~metric:~~} G_{ab}-G_{ak}(G^{-1})_{kl}G_{lb}\, .
\ee
Geometrically, a $D$-dimensional space whose metric is independent of
$d$ coordinates $\th^i$ can be thought of as a bundle $M$ with fiber
coordinates $\th$.  The metric on the fiber is $G_{ij}$, and hence we
refer to $\hat\f$ as the ``invariant fiber dilaton''.  The induced
metric on the quotient space $M/\{\th^i\}$ is the quotient metric
(\ref{quot}).

To prove the invariance of (\ref{lem}), we consider the action of the
generators of $O(d,d,\Z )$ separately.  The $Gl(d,\Z )$ and $\Theta
(\Z )$ transformations trivially leave $\f$ and $det(G_{ij})$
invariant, and hence $\hat\f$ as well.  To show invariance under
factorized duality, we write $\hat\f$ explicitly in terms of $\S$:
\be
\label{fres}
\eqalign{
\hat\f\, =\, &\F+\ln det(I+\S )+\fr12\ln det\fr12\[(I-\S )(I+\S )^{-1}
+(I+\S^t)^{-1}(I-\S^t)\]\cr =\, &\F+\fr12 \ln det(I-\S^t\S ) \, .}
\ee
To get the second equality, we need to split $\ln det(I+\S )$ as
$\fr12\ln det(I+\S ) +\fr12\ln det(I+\S ^t)$.  Under $\S\ra O_2\S
O_1^t$ (\ref{OSO}), $(I-\S^t\S )\ra O_1(I-\S^t\S )O_1^t$, and hence
$\hat \f$ is invariant.  This completes the proof of the invariance of
the fiber dilaton $\hat\f$.

The proof of the invariance of (\ref{quot}) is entirely parallel;
again, the $Gl(d,\Z )$ and $\Theta (\Z )$ transformations trivially
leave (\ref{quot}) unchanged.  To prove invariance under factorized
duality, we express (\ref{quot}) explicitly in terms of $\S ,
\G^1,\G^2,\G$ (see \ref{ij},\ref{F}):
\be
\label{abmess}
\eqalign{
G_{ab}&-G_{ak}(G^{-1})_{kl}G_{lb}\cr
=&\[\fr12(\G+\G^t)-\fr14\{\G^1(I+\S )^{-1}\G^2+
\G^{2t}(I+\S^t )^{-1}\G^{1t}\}\cr
&\,\, +\fr12\{\G^1(I+\S )^{-1}+\G^{2t}(I+\S^t)^{-1}\}
\{(I-\S )(I+\S )^{-1}+(I+\S^t)^{-1}(I-\S^t)\}^{-1}\cr
&\qquad\qquad\qquad\qquad\qquad\qquad\qquad\quad\times
\{(I+\S )^{-1}\G^2+(I+\S^t)^{-1}\G^{1t}\}\]_{ab}\cr
=& \[\fr12(\G+\G^t)+\fr14\{\G^1(I-\S^t\S )^{-1}(\G^{1t}+\S^t\G^2)
+\G^{2t}(I-\S\S^t)^{-1}(\G^2+\S\G^{1t})\}\]_{ab}\, .
\cr}
\ee
The final expression is manifestly invariant under (\ref{OSO}), which
completes the proof of the invariance of the quotient metric
(\ref{quot}).

The theorem (\ref{them}) that $\du\f$ is invariant follows from the
invariance of (\ref{lem},\ref{quot}) together with the identity
(\ref{det}).  This is compatible with results in low-energy effective
field theories \cite{gpo,ven} and string field theory \cite{sen,kz},
and physically implies that the string coupling constant
$g_{string}^{-1}=<e^{\du\f}>=<\sqrt{det{\cal G}}\, e^\f >$ is
invariant under $O(d,d,\Z )$.  The invariance of
(\ref{lem},\ref{quot}) actually prove that $\sqrt{det(G_{ij})}\, e^\f$
and the quotient metric (\ref{quot}) are separately invariant; of
course, this holds only for $O(d,d,\Z )$, and not the full $O(D,D,\Z
)$ that it is embedded in.

\newpage

\newsection{The Heterotic String}

\noindent
In this section, we make a conjecture about the discrete symmetries of
the heterotic string by requiring compatibility with the flat limit
\cite{grv,sw} and with the bosonic case.  We start with a {\it
curved\/} heterotic background, which we assume is a consistent,
conformally invariant, heterotic string theory, with an action:
\be
\label{het}
\eqalign{
S_{het}=&\ind\[{\cal E}_{IJ}(x)\pa X^I\pab X^J+{\cal A}_{IA}(x)\pa
X^I\pab Y^A \cr &\qquad\qquad + E_{AB}\pa Y^A\pab Y^B -\fr14\f
(x)R^{(2)} +({\rm fermionic~terms})\]\, ,}
\ee
and with the second-class constraints that the $Y^A$ are chiral
bosons: $\pa Y^A=0$. As before, $\{X^I\}=\{\th^i,x^a\}$, with
$i=1\dots d$, and $a=d+1\dots D$.  In addition, we have $d_{int}$
internal chiral bosons $Y^A$: $A=1\dots d_{int}$.  In flat space, we
have $d_{int}=16$, but more generally, we may find other solutions
\cite{pol,mny}.  The spacetime background is given by ($\cal E$,$\f$),
as in the bosonic case (\ref{calG},\ref{f}), and, in addition, the
gauge field $\cal A$.  We assume that this curved background is
independent of the $d$ coordinates $\th^i$.  The constant internal
background is
\be
\label{GAB}
E_{AB}=G_{AB}+B_{AB}\, ,
\ee
where $G_{AB}$ is the metric on the internal lattice (one half the
Cartan matrix of the internal symmetry group when the lattice is the
root lattice of a group) and $B_{AB}$ is its antisymmetrization
\cite{egrs}, \ie $E_{AB}$ is upper triangular.  In the spacetime
supersymmetric flat case, the symmetry group is $E_8\times
E_8$.\footnote{The $Spin(32)/Z_2$ string can be described as the
$E_8\times E_8$ string with a particular gauge field background
\cite{ginnsw}.} In curved space, this group is in general different
\cite{pol,mny}.

Following \cite{grv}, the expected symmetry group is isomorphic to
$O(d,d+d_{int},\Z )\subset O(D,D+d_{int},\Z )$ as in section 4, acting
by fractional linear transformations on the $(D+d_{int})\times
(D+d_{int})$ dimensional matrix
\be
\label{xi}
\Xi (x)=\left(\matrix{
{\cal E}_{IJ}+\fr14{\cal A}_{IA}(G^{-1})_{AB} {\cal A}_{JB}&{\cal
A}_{IA}
\cr0&E_{AB}
\cr}\right)
\ee
where $G^{-1}$ is the inverse of $G_{AB}$ in (\ref{GAB}).  The matrix
$\Xi$ is the embedding of the heterotic background into a bosonic
$(D+d_{int})$-dimensional background, and the group $O(D,D+d_{int})$
is the subgroup of $O(D+d_{int},D+d_{int})$ that preserves the form
(\ref{xi}).  Note that the spacetime metric ${\cal G}_{IJ}$ (the
symmetric part of ${\cal E}_{IJ}$ in (\ref{xi})) is the quotient
metric of the $(D+d_{int})$-dimensional space modulo $\{ Y^A\}$ (here,
$G_{AB}$ is the fiber metric).  This leads to a simple expression for
the transformation of the dilaton:
\be
\label{hetdil}
\f'=\f+\fr12\ln\Big( \fr{det{\cal G}}{det{\cal G'}}\Big)\, ;
\ee
note that this is independent of the gauge fields.

\newsection{Applications}

\noindent
In this section we explore a number of consequences of the discrete
symmetries.  We first discuss an exact $D=3$ closed string background
that is independent of $d=2$ coordinates \cite{ils}.  We then turn to
$D=2$ heterotic backgrounds
\cite{pol,mny}.  In both cases, we find that uncharged black compact
objects (strings or holes) are equivalent to charged $D=2$ black
holes.\footnote{While writing up our results, we found similar
observations in \cite{hhs}.}

\newsubsection{The closed string example}

\noindent
The simplest nontrivial example after the $D=2$ black hole duality
\cite{gdvv} is
a compact black string given by attaching a circle to every point of
the $D=2$ black hole spacetime ($Sl(2,\R )_k/U(1)\times U(1)$).  To
leading order, the action is
\be
\label{bstr}
S_{Black String}=\ind \[k(\pa x \pab x + \t\pa\th^1\pab\th^1)
+\a\pa\th^2\pab\th^2 -\fr14\f (x)R^{(2)}\]\, ,
\ee
where
\be
\f (x)=\f_0 + \ln (\ch)\, .
\ee
The first term in $S_{Black String}$ is the euclidean black hole
metric \cite{bh}, and the second term describes a circle of radius
$\sqrt\a$ attached to each point.  This $D=3$ background is
independent of $d=2$ coordinates $\th^i$, and is described by the
matrix (\ref{calG})
\be
\label{back}
{\cal E} =\left(\matrix{E&0\cr 0&F\cr}\right)\, ,
\ee
where
\be
\label{start}
E=\left(\matrix{k\, \t&0\cr 0&\a}\right)\q F=k \, .
\ee

The group of generalized duality tranformations $O(2,2,\Z )$ maps this
background into other backgrounds that (in general) have different
spacetime interpretations.  {Since} $F_1=F_2=0$ in (\ref{back}, \cf
\ref{calG}), $O(2,2)$ acts on the background by transforming only $E$
and $\f$ as given in (\ref{tE}) and (\ref{tdil}). A particularly
interesting point on the trajectory of $O(2,2,\Z )$ is reached by
acting with the element
\be
g=\mat I\Theta 0I\mat p00p\mat 0II0 \mat I{-\Theta}0I =\mat{\mat
100{-1}}{\mat 0000}{\mat 0110}{\mat 100{-1}}\, ,
\ee
where
\be
I=\mat 1001 \q \Theta=\mat 01{-1}0 \q p=\mat 0110 \, .
\ee
This transforms $E$ and $\f$ to
\be
\eqalign{
E'=g(E)&=\fr1{1+\a k\, \t}\mat {k\,\t}{\a k\,\t}{-\a k\, \t}{\a} \cr
&\cr &=
\fr1{\ch -\l}\mat {k(1-\l )\sh}{\l\sh}{-\l\sh}{\l\ch /k}\cr}
\ee
\be
\label{dualdil}
\f'(x)=\f_0-\ln (1-\l ) +\ln (\ch -\l ) \, ,
\ee
where
\be
\label{lam}
\l=\fr{k\a}{1+k\a} \, .
\ee
This gives an action
\be
\label{charge}
\eqalign{
S_{Charge} &= \ind \[k(\pa x \pab x + \fr{(1-\l )\sh}{\ch-\l}
\pa\th^1\pab\th^1)\cr
&\cr
\qquad\qquad&+
\fr{\l\sh}{\ch-\l}(\pa\th^1\pab\th^2-\pa\th^2\pab\th^1)
+\fr{\l\ch}{k(\ch-\l )}\pa\th^2\pab\th^2 -\fr14\f'(x)R^{(2)}\]\, ,\cr}
\ee
and, after wick rotating $\th^1\ra it$, corresponds to a charged black
hole of the type found in \cite{ils} with mass $M$
\be
\label{mass}
M=(1-\l )M_0=\sqrt{\fr2k}e^{\f_0}\q M_0=\sqrt{\fr2k}e^{\f'_0}\, ,
\ee
and charge
\be
\label{q}
{\cal Q}=\sqrt{\l (1-\l )}\fr{2M_0}k=\sqrt{\fr\l{1-\l}}\fr{2M}k \, ,
\ee
where $\f'_0=\f_0-\ln (1-\l )$ is the constant part of the dual
dilaton (\ref{dualdil}).  The action (\ref{charge}) is related to the
precise action for the coset $(Sl(2,\R )_k\times U(1))/U(1))$
\cite{ils} by rescaling $\th^2\ra k\th^2$.\footnote{Our $k$ matches
\cite{bh}, which is $2k$ of \cite{ils}.}

This shows that the charged black hole is equivalent to the compact
black string as a conformal field theory.  We now consider some
particular limits of this solution.

The limits $\a\ra 0,\infty$ in (\ref{bstr}) is the $2D$ black hole
$\times$ a degenerate circle \cite{bh}, with the two limits related by
$R\ra 1/R$ duality.  These limits correspond to the limits $\l\ra 0,1$
in (\ref{lam}).  In $S_{Charge}$ (\ref{charge}), $\l\ra 0$ is
precisely the $2D$ black hole $\times$ the same degenerate circle;
however, the $\l\ra 1$ limit gives the action (modulo an integer total
derivative term)
\be
S_{\l\ra 1}=\ind \[k\pa x \pab x + \fr1kcoth^2x\pa\th^2\pab\th^2
+(1-\l )\pa\th^1\pab\th^1 -\fr14\f' (x)R^{(2)}\]\, ,
\ee
$$
\f' (x) = \f_0-\ln (1-\l ) + \ln(\sh ) \, ,
$$

\noindent
which corresponds to the dual $2D$ black hole $\times$ a degenerate
circle \cite{gdvv}.  In both cases, the degenerate limits are
equivalent as CFT's to a {\it noncompact\/} black string.

\newsubsection{The heterotic string example}

\noindent
We focus on the example of \cite{mny}, which is a $D=2$ heterotic
string with internal degrees of freedom taking values in a standard
12-dimensional lattice (the vector weights of $SO(24)$).\footnote{More
precisely, only spacetime bosons have internal quantum numbers in the
vector representations of $SO(24)$; spacetime fermions have internal
quantum numbers in the spinor representations of $SO(24)$.} We find
that a family of charged black holes (and naked singularities) are
dual to a neutral one, which is the exact CFT given by the heterotic
$D=2$ black hole \cite{mny}.

We start with a heterotic $D=2$ action:
\be
\label{shet}
\eqalign{
S_{het}=\ind \[ k(\pa x\pab x +\t\pa\th\pab\th) +\pa Y^A\pab Y^A
-\fr14\f (x) R^{(2)}&\cr + \rm{(fermionic~terms)}&\]\, ,\cr}
\ee
where $A=1,\dots ,12$, $k=5/2$ (for criticality), and $\f=\f_0+\ln
(\ch)$.  This action describes a neutral heterotic $D=2$ black hole.
The conformal field theory (\ref{shet}) corresponds to a background
(\ref{xi})
\be
\label{bhet}
\Xi=\left(\matrix{k&0&0\cr 0&k\,\t&\cr 0&0&I\cr}\right)\, ,
\ee

\noindent
where the internal background $I$ is the $12\times 12$ identity matrix
corresponding to the vector weights of $SO(24)$.  Only a $2\times2$
block $E$ of the matrix $\Xi$ is affected by the discrete
transformations we discuss in this example:
\be
\label{toy}
E=\mat{k\,\t}001\, .
\ee

By transforming $E$ and $\f$ with a group element $g_n\in O(1,2,\Z )
\subset O(1,13,\Z )$ (where $n $ is an arbitrary integer):
\be
\label{gggg}
g_n=\mat0II0\mat I{n\Theta}0I\mat{A_n^t}00{A_n^{-1}}=
\mat{\mat0000}{\mat10{-n}1}{\mat1n01}{\mat{-n^2}n{-n}0}\, ,
\ee
where
\be
I=\mat1001\q \Theta=\mat01{-1}0\q A_n=\mat10n1 \, ,
\ee
one finds
\be
\label{Ehet}
g_n(E)=E_n'=\mat{(n^2+k\,\t )^{-1}}{-2n(n^2+k\,\t )^{-1}}01\, ,
\ee
\be
\label{dhet}
\f' (x)=\f_0+\ln (n^2+k)+\ln (\ch -\fr k{n^2+k}) \, .
\ee

After rescaling
\be
\th\ra \fr{k+n^2}{\sqrt{k}}t\, ,
\ee
and defining $r$ to be a linear function of the dilaton $\f'$
(\ref{dhet}),
\be
\label{rho}
Qr = \ln (\ch -\fr k{n^2+k}) \, ,
\ee
where $Q$ is a constant determined below, the background (\ref{Ehet})
gives rise to an action\footnote{Recall that ${\cal
G}_{tt}=\Xi_{tt}'-\fr14 {\cal A}^2$ (see \ref{xi}).}
\be
\label{Lchar}
\eqalign{
S_{Charge}=\ind \[ f(r )\pa t \pab t + f(r )^{-1} \pa r\pab r -A(r)\pa
t \pab Y^1 + \pa Y^A\pab Y^A&\cr -\fr14\f' (r) R^{(2)} + {\rm
(fermionic~terms)}&\]\, , }
\ee
with
\be
f(r)=1-2me^{-Qr}-q^2e^{-2Qr},
\ee
\vskip .1in
$$ A(r)=nQ+2qe^{-Qr} \q \f' (r)=Qr+\f_0+\ln (n^2+k)\, , $$
\vskip .1in
\noindent
where $Q=2/{\sqrt{k'}}$ is determined by the normalization of ${\cal
G}_{rr}$ in (\ref{Lchar}), and
\be
\label{QMq}
 2m=\fr{n^2-k'}{n^2+k'} \q q=\fr{n\sqrt{k'}}{n^2+k'}\, .
\ee
Following~\cite{bh} we have replaced $k$ with $k'=k-2=1/2$ in
(\ref{QMq}).  Wick rotating $t\ra it$, along with $q\ra -iq$
(necessary to maintain hermiticity of the action), the theory
(\ref{Lchar} with $|n|>1$) describes a $D=2$ charged black hole with
mass and charge \cite{mny}
\be
M=Q(n^2-k')e^{\f_0}\q {\cal Q}=n\sqrt{8}e^{\f_0}\, .
\ee
For $n=-1,0,1$, the theory (\ref{Lchar}) describes a naked
singularity.

We emphasize that these backgrounds, for all $n$, are different
spacetime interpretations of the {same\/} CFT: the exact CFT given by
the neutral heterotic $D=2$ black hole.

\newsection{Concluding Remarks and Open Problems}

\noindent
We have shown that $O(d,d,\Z )$ acts on the space of backgrounds that
are independent of $d$ coordinates.  We expect that in general the
full symmetry group acting on the space of $D$-dimensional curved
backgrounds is larger.  Some of these extra symmetry generators can be
found by considering quotients of $(D+d)$-dimensional actions that are
more general than (\ref{Dd}); we hope to discuss this somewhere else.
Ideally, one would like to find the complete symmetry group for the
space of all curved backgrounds.

The $O(d,d,\Z )$ subgroup already leads to interesting relations
between different geometries. We have illustrated this with charged
black hole examples; similar studies in the context of string
cosmology may lead to surprising consequences.

Elements of $O(d,d,\Z )$ with $det = -1$ relate backgrounds with
(possibly) different topologies.  In the flat case, such
transformations coincide
\cite{gs}
with mirror symmetry for $N=2$ superconformal backgrounds
\cite{gpcan}.  It would be interesting to understand the relation
between the two in the general case.

Another open problem is the issue of higher order corrections; this is
a problem for the spacetime interpretation of quotients as well as for
discrete symmetry transformations.

\vspace{1cm}
\noindent
{\sc Acknowledgments:} We would like to thank R. Dijkgraaf, C. Nappi,
E. Verlinde, E. Witten, and B. Zwiebach for discussions. The work of
AG is supported in part by DOE grant No. DE-FG02-90ER40542, and that
of MR is supported by the John Simon Guggenheim Foundation.

\vspace{1cm}

\font\twelvebf=cmbx12
\newcommand{\savebold}{\fam\bffam\twelvebf}
{\renewcommand{\Large}{\normalsize}
\renewcommand{\bf}{\sc}
\newcommand{\np}{Nucl.\ Phys.\ }
\newcommand{\pr}{Phys.\ Rev.\ }
\newcommand{\cmp}{Commun.\ Math.\ Phys.\ }
\newcommand{\pl}{Phys.\ Lett.\ }

}

\end{document}